# Lung Imaging with UTE MRI

**Peder E. Z. Larson**

Department of Radiology and Biomedical Imaging, University of California, San Francisco, CA, 94143, USA

Corresponding Author: Peder Larson (Peder.Larson@ucsf.edu)

**Index**

## 1 INTRODUCTION

### 1.1 Challenges and Opportunities for MRI of the Lungs

Cross-sectional imaging of the lungs, or pulmonary imaging, has proven to be an incredibly valuable tool in a wide range of pulmonary diseases. The vast majority of lung imaging is done with CT, as it is fast enough to freeze respiratory motion and provides high spatial resolution to visualize fine structure of the lungs (e.g., airways, blood vessels, and lung parenchyma).

MRI of the lungs is inherently challenging due to the presence of large local magnetic field gradients, relatively low proton density, and motion. The lungs consist of airways, blood vessels and parenchyma that includes microscopic air sacs called alveoli. There is a relatively large difference in magnetic susceptibility between air and lung tissue, which leads to large gradients in the magnetic field within the lung. This, in turn, leads to relatively short transverse relaxation times, $T_2$ and $T_2^*$. $T_2^*$ is particularly short ($T_2^* = 0.8$ ms at 3T and 2.1 ms at 1.5 T [1]) because of the intravoxel dephasing caused by the large tissue gradients in the magnetic field. $T_2^*$ lengthens significantly at lower field strengths ($T_2^* = 8$-$10$ ms at 0.55 T [2,3]), an emerging opportunity discussed in this chapter. The proton density in lung parenchyma is also much lower than that of other soft tissues, as much of the parenchymal space is filled with air. Furthermore, the lungs are always moving, making motion a key challenge in performing lung MRI.

The benefits of performing MRI for lung imaging include no ionizing radiation, opportunities for multiple contrasts, and integration with other MRI scans. MRI requires no ionizing radiation compared to CT and PET/SPECT, where exposure increases the lifetime risk of cancer [4]. This makes MRI desirable in populations of patients such as those in pediatrics and obstetrics where radiation sensitivity is a particular issue. This is especially true when repeated lung imaging scans are required. MRI also offers the opportunity to obtain multiple tissue contrasts. The most common lung MRI techniques are structural $T_1$-weighted scans, but also emerging are functional contrasts such as ventilation and perfusion, as well as other MRI contrast mechanisms including $T_2$-weighting and diffusion-weighting. Finally, lung MRI can be combined with other MRI scanning techniques, including cardiac MRI, abdominal MRI, whole-body MRI, and PET/MRI, for increasing examination efficiency by only requiring a single scan session and providing more comprehensive assessment that includes evaluation of the pulmonary system.

## 1.2     Why use UTE for MRI of the Lungs?

Ultrashort echo time (UTE) MRI has emerged as the leading approach for lung MRI due to two key advantages. First, and arguably the most significant advantage, is that it can efficiently capture rapidly decaying short-$T_2^*$ signals from lung tissue. The other advantage of UTE MRI is motion management. It is inherently less likely to produce motion artifacts, it includes information for motion tracking, and it is also extremely well suited to motion compensated and motion corrected reconstructions. This advantage comes from using center-out k-space trajectories, where repeated measurements of the center of k-space can be used both to monitor motion and to alleviate potential artifacts.

## 2  METHODS FOR UTE LUNG MRI

This section describes 3D UTE pulse sequences, however, other pulse sequences, such as ZTE, 2D UTE, and other gradient echo sequences can be used as well.

### 2.1  Pulse sequences

One of the most successful UTE pulse sequences is shown in Fig. 1. It is a 3D sequence consisting of a slab selective excitation and a center-out 3D k-space trajectory [5]. The slab selection reduces spatial encoding requirements and artifacts originating outside of the lungs. Because of its short refocusing gradient, it does not introduce any significant increase in TE. 2D UTE sequences are less common because they provide limited spatial coverage and the half-pulse excitations they require are extremely sensitive to system delays and eddy currents.

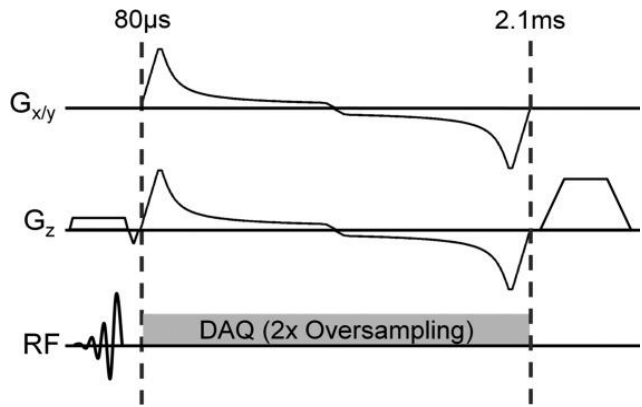

Figure 1. Example 3D UTE pulse sequence for lung MRI featuring slab excitation and a variable density arc-length optimized out-and-back radial readout. (Reproduced with permission from Ref. [5])

### 2.2  k-space Trajectory

A large group of 3D non-Cartesian k-space trajectories has been successfully used for UTE lung MRI, with their most important features being repeated center of k-space sampling, the ability to undersample for accelerated scanning, and the ability to perform pseudo-random temporal ordering for retrospective motion correction (Fig. 2). 3D radial trajectories, also known as kooshball trajectories or projection-reconstruction, are the simplest trajectories. They also cover the smallest k-space area per TR and suffer from reduced signal-to-noise ratio (SNR) efficiency due to their high sampling density at the center of k-space, although this concern can be greatly alleviated with variable density readouts [5]. 3D spiral trajectories such as twisted projections [6], cones [7], and FLORET (Fermat looped, orthogonally encoded trajectories) [8] cover greater areas of k-space per TR and can be flexibly designed for given readout durations and undersampling. The radial cones trajectory has the advantage of providing excellent control of undersampling [9]. Stack-of-stars and stack-of-spirals trajectories also provide efficient 3D coverage [10], but Cartesian encoding creates increased likelihood of motion artifacts in the stack dimension.

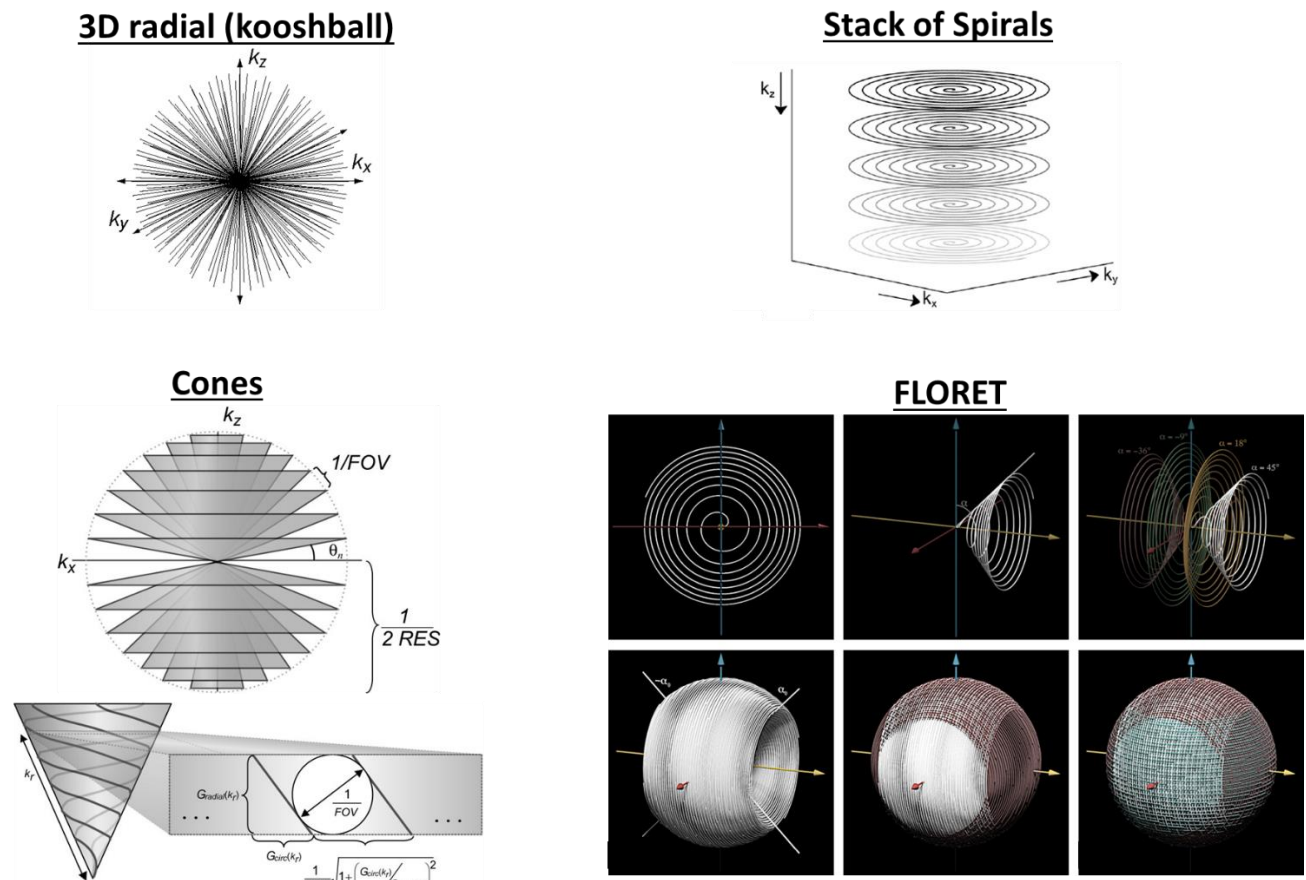

Figure 2. 3D center-out k-space trajectories that have been the most successful for UTE Lung MRI, including 3D radial or kooshball sampling, 3D cones [7], stack-of-spirals [11], and FLORET [12]. (Adapted with permission from Refs. [7, 11, 12])

## 2.3 Motion Management

Strategies for motion management can be approximately divided into prospective methods such as breath-holding or gating, and retrospective methods that sort data based on some estimation of respiratory motion.

Breath-holding is effective and clinically common, but must be completed in 10-20 seconds which limits the achievable spatial resolution and coverage.

Free-breathing scanning with prospective gating aims to acquire data only during the quiescent phase of respiration, and allows increased spatial resolution and coverage compared to breath-holding. It relies on realtime measurement of respiratory motion (e.g., bellows belt, navigators, pilot tone) to trigger data acquisition. The main disadvantages are that it is not continuously acquiring data which reduces SNR efficiency and any irregular motion (e.g., bulk motion, coughing, shallow vs. deep breathing) may lead to additional artifacts.

Free-breathing with retrospective gating uses continuous data acquisition, and upon completion assigns data to bins for reconstruction. Using continuous acquisitions allows reconstruction of multiple respiratory phases. Continuous acquisitions can also improve SNR efficiency. In addition, retrospective motion estimations can be more accurate than prospective methods particularly for irregular motion. It requires sufficiently dispersed temporal ordering of the sampling (e.g., golden angle type methods) for retrospective binning of data based on estimated respiratory motion. When a subset of the data is retrospectively binned based on the respiratory motion, it should be relatively evenly distributed in k-space.

The methods for estimation of motion include: DC (k-space center signal) navigators – here the repeated center of k-space (k=0) signal is used, typically with some signal processing, to estimate respiratory motion over time [13]; 1D navigators – these are most common in stack-of-stars or stack-of-spirals, where applying 1D Fourier Transform (FT) to data in the stack dimension provides a 1D image that can be processed to estimate respiratory motion [14, 15]; image-based navigators – these methods use a larger area around the center of k-space to produce low-resolution dynamic images from which motion can be estimated. This has a distinct advantage in depicting different motion patterns and bulk motion [16–18] (Fig. 3).

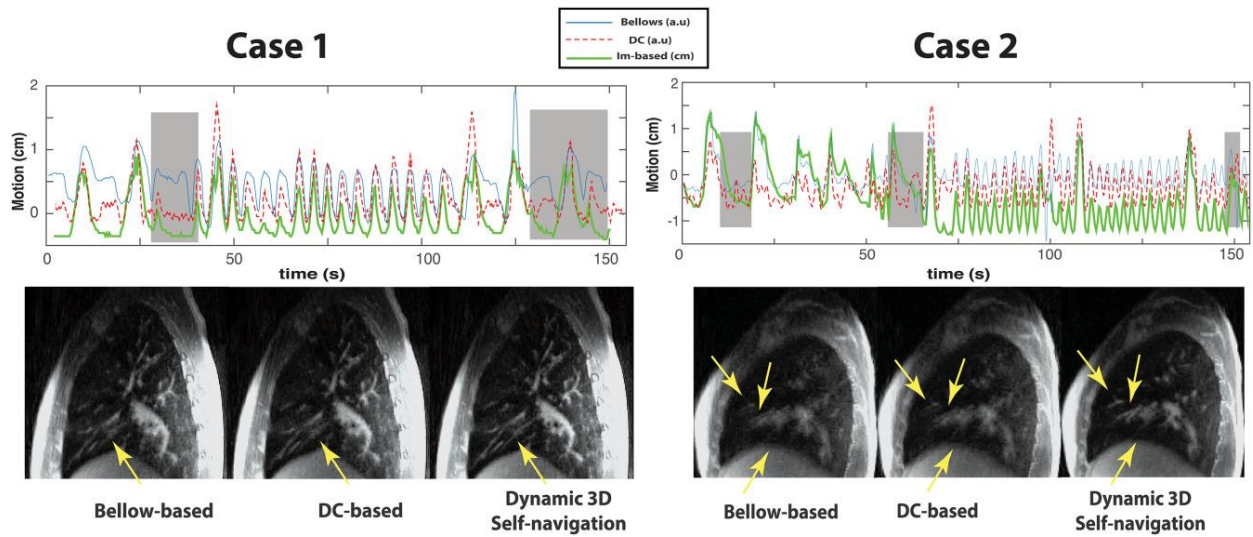

Figure 3. Comparison of bellows belt, DC (k-space center) and image-based navigation for UTE Lung MRI. Case 1 is a cystic fibrosis patient with mildly irregular breathing, while Case 2 is a cystic fibrosis patient with highly irregular breathing. All motion management methods perform similarly in Case 1, but the ability of image-based navigators to capture the more variable and complex motion in Case 2 leads to significant improvement in image quality as shown by the arrows. (Reproduced with permission from Ref. [16])

## 2.4 Image Reconstruction

The 3D non-Cartesian k-space trajectories used for UTE MRI require non-Cartesian image reconstruction methods. The most straightforward and fastest methods are gridding and the use of non-uniform fast FT (FFT), but the current state-of-the-art is based on iterative reconstructions that include capabilities for parallel imaging and compressed sensing acceleration. 3D non-Cartesian trajectories are well suited to these types of acceleration as they can be uniformly or pseudo-randomly undersampled.

Advanced iterative image reconstructions can also improve the image quality, motion management, and the information provided by UTE lung MRI, and are typically used with free-breathing, and retrospectively gated acquisitions. One approach to improve image quality is to use motion-compensated reconstructions [19, 20]. These require data to be binned and images reconstructed for different respiratory states. These different images are then aligned using deformable image registration, allowing the data from all respiratory states to be combined into a single image. By using data from the entire scan time, these approaches are more SNR efficient and thus can provide improved SNR and resolution.

Free-breathing data can also be reconstructed into multiple respiratory states, creating motion resolved images that improve motion management and also provide dynamic images of respiration that can be used to measure tissue motion and ventilation. The quality of motion resolved reconstructions is improved by jointly reconstructing all respiratory state images. This can be done with a sparsity or low-rank penalty across the state dimension in an iterative reconstruction [14]. These are often referred to as eXtra-Dimensional or XD reconstructions (Fig. 4).

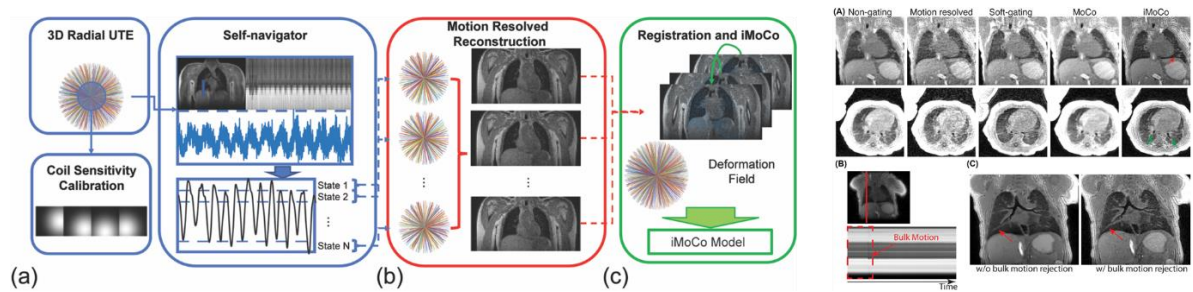

Figure 4. Illustration of advanced image reconstruction methods for UTE lung MRI and sample results. The depicted technique is iterative motion compensation (iMoCo) reconstruction, which includes a motion state identification with self-navigation, a motion-resolved (XD) reconstruction, and a motion compensated reconstruction [(a) to (c)]. Results in a 10 week old infant (right) show progressive improvements in image quality, where iMoCo efficiently takes advantage of all the acquired data. Note in this example a dynamic image navigator identified a period of bulk motion that could be removed to further improve image quality. (Reproduced with permission from Ref. [19])

## 3 CONTRAST MECHANISMS WITH UTE MRI

### 3.1 $T_1$-weighting

3D UTE MRI nominally provides $T_1$-weighted contrast, as it uses short TRs and typically flip angles tuned to the optimal Ernst angle for lung parenchymal signal. This provides clear delineation of parenchyma, vessels, and airways within the lungs as well as well defined borders between surrounding tissue and other organs.

### 3.2 Non-contrast Ventilation

Ventilation refers to the circulation and exchange of gasses in the lungs, and is a critical component of lung function. Ventilation can be assessed by measuring the change in lung tissue density or volume during respiration, and this forms the basis of non-contrast ventilation imaging (Fig. 5) [14, 21, 22].

With UTE MRI, non-contrast ventilation can be measured from motion-resolved reconstructed images, where the respiratory states are aligned using deformable image registration. After alignment, the change in signal amplitude can be measured as so-called "specific ventilation," and regional changes in volume can be measured as so-called "regional ventilation."

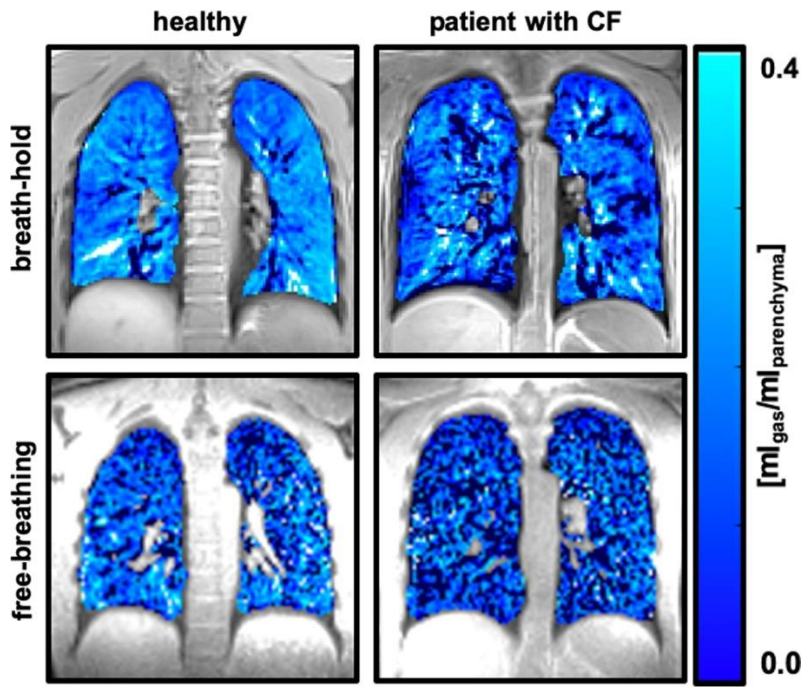

Figure 5. Example of UTE lung ventilation maps, showing more inhomogeneous ventilation patterns in the cystic fibrosis (CF) subject on the right. In this work, similar results were found whether ventilation was calculated based on signal intensity changes or tissue volume changes. (Reproduced with permission from Ref. [22])

## 3.3 Oxygen Enhanced Ventilation

Pure oxygen is paramagnetic whereas most tissue is diamagnetic. Due to this magnetic susceptibility difference, breathing in 100% $O_2$ results in a shortening of $T_1$ in the lung parenchyma compared to breathing room air [23–25]. This change is easily observed with typical $T_1$-weighted 3D UTE sequences. Ventilation can then be measured by comparing $T_1$-weighted images or $T_1$ values between breathing room air and 100% $O_2$ (Fig. 6).

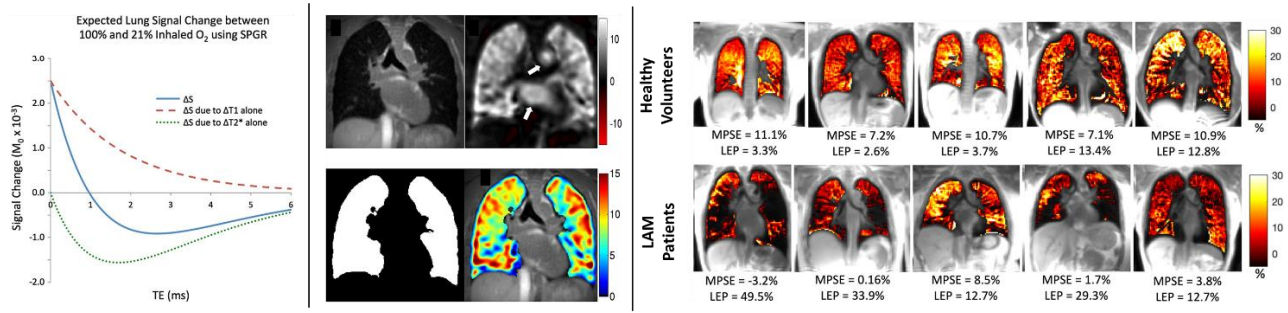

Figure 6. Oxygen-enhanced ventilation imaging. (left) With UTE, theory predicts that the signal increase between room air and 100% $O_2$ is due to $T_1$ shortening without sensitivity to signal loss due to $T_2^*$ shortening effects. (middle) UTE datasets from breathing room air and 100% $O_2$ are subtracted to show the ventilation effect, which can be mapped onto the lung anatomy but oxygen enhancement in the aorta and ventricles (arrows) is also visible. (right) Examples of UTE oxygen-enhanced ventilation maps in healthy volunteers and lymphangioleiomyomatosis (LAM) patients acquired at 0.55 T. (Reproduced with permission from Ref. [23, 26])

## 3.4 New Frontiers

Recently, mid/low field (< 1 T) MRI scanners have been revisited for cost-efficiency benefits. Lower field strengths are advantageous for lung MRI because $T_2^*$ is increased due to the reduced effect of magnetic susceptibility differences between air and lung tissue (e.g., $T_2^* = 8\text{-}10$ ms at 0.55 T [2, 3]). This allows longer readouts and more SNR efficient acquisitions that can offset some of the losses due to reduced polarization at lower magnetic fields [27]. It also opens up a greater range of contrasts and reduces the need for UTE pulse sequences.

Zero echo time (ZTE) pulse sequences have also been applied for lung MRI as they can also efficiently capture rapidly decaying signals [28]. However, they have typically not performed as well as UTE MRI. One reason is the lack of slab selection, which in turn requires encoding of much larger FOVs and also increased susceptibility to motion artifacts from abdominal organs. ZTE MRI also has a more limited range of available flip angles than UTE, especially over larger FOVs [29]. This also limits the SNR efficiency of ZTE.

Ventilation and perfusion measurements in the lung have been achieved using fast 2D scanning, with techniques such as Fourier Decomposition [30] and phase resolved functional lung (PREFUL) imaging [31]. Ventilation is measured using the non-contrast ventilation strategies described above. Perfusion imaging is achieved by binning the images by cardiac state and measuring changes in lung signal intensity. Using 2D scanning creates time-of-flight enhancement of inflowing spins and thus perfusion contrast. These approaches can be used with fast 2D gradient-echo pulse sequences.

# 4    CONCLUSION

MRI has advantages compared to CT for cross-sectional lung imaging of no ionizing radiation, a broader range of contrast, and integration with MRI of other tissues. UTE has greatly advanced lung MRI because it provides increased SNR for the short-$T_2^*$ lung parenchyma and is well-suited to manage motion. Recent advances in lung MRI include ventilation and perfusion mapping, and the use of < 1 T MRI scanners where $T_2^*$ is much longer.

# 5    RESOURCES

We have gathered software resources for performing lung MRI in the following GitHub Organization: https://github.com/PulmonaryMRI/ which includes largely includes image reconstruction methods. All major MRI vendors now provide UTE and/or ZTE MRI pulse sequences for lung MRI, primarily as works-in-progress packages, enabling more research in this area.